# Functional analysis of the N-terminal basic motif of a eukaryotic satellite RNA virus capsid protein in replication and packaging

**Venkatesh Sivanandam[1], Deborah Mathews[1], Rees Garmann[2#], Gonca Erdemci-Tandogan[3], Roya Zandi[3] and A. L. N. Rao[1]***

[1]Department of Plant Pathology & Microbiology, University of California, Riverside, CA 92521, USA

[2]Department of Chemistry & Biochemistry, University of California, Los Angeles, CA, 90095, USA

[3]Department of Physics & Astronomy, University of California, Riverside, CA 92521, USA

*Corresponding author E-mail address: arao@ucr.edu

[#]Present Address: Harvard John A. Paulson School of Engineering and Applied Sciences, Harvard University, Cambridge, MA 02138

# ABSTRACT

Efficient replication and assembly of virus particles are integral to the establishment of infection. In addition to the primary role of the capsid protein (CP) in encapsidating the RNA progeny, experimental evidence on positive sense single-stranded RNA viruses suggest that CP also regulates RNA synthesis. Here, we demonstrate that replication of Satellite tobacco mosaic virus (STMV) is controlled by the cooperative interaction between STMV capsid protein (CP) and the helper virus (HV) Tobacco mosaic virus (TMV) replicase. We identified that the STMV CP-HV replicase interaction requires a positively charged residue at the third position (3R) in the N-terminal 13aa motif. Far-Northwestern blotting showed that STMV CP promotes binding between HV-replicase and STMV RNA. An STMV CP variant having an arginine to alanine substitution at position 3 in the N-terminal 13aa motif abolished replicase-CP binding. The N-terminal 13aa motif of the CP bearing alanine substitutions for positively charged residues located at positions 5, 7, 10 and 11 are defective in packaging full-length STMV, but can package a truncated STMV RNA lacking the 3' terminal 150 nt region. These findings provide insights into the mechanism underlying the regulation of STMV replication and packaging.

## Introduction

*Satellite tobacco mosaic virus* (STMV) was originally found in association with helper virus (HV) *Tobacco mosaic virus* (TMV strain U5) in natural infections of tree tobacco (*Nicotiana glauca*) [1]. The replication of STMV depends on the RNA-dependent RNA polymerase (RdRp) encoded by TMV U5, which has since been renamed *Tobacco mild green mosaic virus*. Other strains of TMV such as U1 and U2 also have been shown to support the replication of STMV [1]. The genome of STMV is composed of a single-stranded, positive-sense RNA of 1059 nucleotides (nt) [2,3]. STMV RNA does not show appreciable sequence homology with helper TMV RNA, except for the 3'-terminal150 nt that exhibit 65% homology with the corresponding tRNA-like structure (TLS) located at the 3' end of TMV RNA [2]. Biologically active icosahedral virions of STMV are 17-18 nm in diameter with a T=1 symmetry [4] and are assembled from 60 identical capsid protein (CP) subunits of 17.5 kDa [2]. The 3D structure of STMV has been determined at 1.8 Å [5].

A hallmark feature of STMV is that it is the only satellite virus, whose HV is a rod-shaped tobamovirus. Additionally, STMV stands out among other spherical viruses because of the high degree of order adopted by the RNA within the capsid[6]. Despite its simple genome organization, information concerning STMV replication is limited[7,8]. Previous studies[7] revealed that a mutation engineered to completely block CP synthesis severely reduced the replication of STMV while a variant designed to express only the N-terminal13 amino acid (aa) motif retained wild type replication via an unknown mechanism.

Using an *Agrobacterium*-based transient expression system (agroinfiltration) we recently reported several interesting properties of STMV when expressed in the presence and absence of its HV RdRp [9]. In the current study, we extended the agroinfiltration system to further shed light on the contribution of the CP, more specifically the N-terminal 13aa motif to STMV replication. We have engineered a series of mutations affecting the positively charged amino acids located within the N-terminal 13aa motif and evaluated their effect on replication and packaging. Application of a wide range of molecular techniques revealed that replication and packaging are regulated by protein-protein (i.e. CP and HV-RdRp) and protein-RNA-protein (i.e. HV-RdRp-STMV RNA-CP) interactions. First, we found that, the N-terminal 13aa motif has a regulatory role in STMV replication. Second, the genomic context of the N-13aa motif has a profound influence in modulating (+) and (-)-strand accumulation. Third, a single positively charged arginine residue located at the third position in the N-terminal 13aa motif of the CP is obligatory for the interaction between CP and HV-RdRp. Fourth, a Far-Northwestern analysis revealed that CP is obligatory to promote interaction between HV-RdRp and STMV RNA (i.e. protein- protein-RNA interaction). Finally, we observed that STMV CP bearing alanine substitutions for positively charged residues located at positions 5, 7, 10 and 11 specifically are defective in packaging full-length STMV genomic RNA. This study shows that packaging and replication of STMV RNA are regulated by the N-terminal 13aa motif of the CP.

# Results

**Evidence showing that the N-terminal 13aa motif of STMV CP is obligatory for replication.** The genome organization of STMV RNA and the sequence of the first 25 N-terminal aa of the CP gene are shown in Fig. 1A. Among these 25 aa, the first 13aa are rich in basic residues and are predicted to interact with RNA inside the assembled capsid [10]. Application of DisProt [11] predicted that the N-terminal 25 aa region is disordered while the remaining 134 aa are highly ordered (Fig. 1B). Preliminary data on the replication of STMV in plant protoplasts suggested a role for STMV CP and more specifically for the N-terminal13aa in replication [7]. We opted to extend the *Agrobacterium*-mediated transient expression system (agroinfiltration) developed recently for STMV in our lab [9] in dissecting the mechanism involving the CP-regulated STMV replication. To this end, we constructed three independent variant agrotransformants of wt STMV (pSTMV; Fig. 1C). These are as follows: (i) Agrotransformant CPKO was constructed by substituting the translation initiating methionine codon located at nucleotide 162-164 with a stop codon (i.e. $^{162}AUG^{164} \rightarrow {}^{162}UAG^{164}$) (Fig. 1C); (ii) agrotransformant CPΔ13aa was designed to express the CP devoid of the first 13aa by engineering a stop codon at the translation initiation site and a start codon ataa14 (Fig. 1C) and (iii) agrotransformant CP13aa was designed to express only the first 13aa by engineering a stop codon at aa position 14 (Fig. 1C).

Each of the above-mentioned three agrotransformants was mixed with agrocultures of pRP to provide the HV RdRp (Fig. 1C) and co-infiltrated into *N. benthamiana* leaves. Plants co-infiltrated with pSTMV and pRP served as positive controls. At five days post infiltration (dpi), total RNA and protein preparations were isolated and subjected respectively to duplicate Northern blot hybridization to detect progeny (+) and (-)-RNA and Western blot analysis. Duplicate Northern blots probed for assessing the affect of CP on the accumulation of progeny (+) and (-)-strands are shown in Fig. 2A. Quantitative analysis of progeny RNA with respect to the wt control is tabulated and shown in Fig. 2B. This experiment was repeated three times and we consistently observed that the presence of CP, more specifically the N-terminal13aa, had profound influence on progeny accumulation. Quantitative data in Fig. 2B represents the relative (+) and (-)-strand accumulation for each CP mutant when compared internally to that of wt. It was observed that, complete absence of CP, exemplified by the behavior of CPKO, decreased the plus-strand accumulation by 90% while minus-strand accumulation was not affected. By contrast, CP lacking the N-terminal 13aa motif (i.e. CPΔ13aa) reduced the plus-strand accumulation by 47% while a 3-fold increase in minus-strand accumulation was observed. Interestingly, expression of only the N-terminal 13aa motif had no detectable effect on plus-strand accumulation while a 4-fold increase in minus-strand accumulation was observed (Fig. 2B). Taken together, these results suggested that the N-terminal13aa region of the STMV CP has two independent roles for regulating (-) and/or (+)-strand synthesis (see Discussion). With respect to CP accumulation, as expected, in addition to the wt control, a detectable level of a faster migrating CP was accumulated for variant Δ13aa but not for other variants (Fig. 2A). To detect the CP of 13aa by Western blot, the 13aa was FLAG tagged and its expression was confirmed by using anti-FLAG monoclonal antibodies (Fig 6A).

Recently, we demonstrated that production of a truncated form of STMV RNA is a hallmark feature associated with STMV replication in *N. benthamiana* [9]. It was observed that with either mechanical inoculation of STMV with its HV or co-expression of STMV with HV-RdRp via agroinfiltration, a truncated non-replicative form of STMV lacking the 3' terminal 150 nt (i.e. STMVΔ150) was also accumulated along with the wt genomic RNA. Additional experiments demonstrated that production of STMVΔ150 is linked to HV-dependent replication and is not a product of ribonuclease activity [9].To verify the production of STMVΔ150 in leaves expressing each of the three variant agrotransformants (CPKO, Δ13aa and 13aa), total RNA preparations were subjected to RT-PCR. Production of STMVΔ150 was associated with either wt STMV (Fig. 2A, lane 1 in RT-PCR panel) or a variant competent to express the 13aa (i.e. CP 13aa in Fig. 2A, lane 5 in RT-PCR panel) but not with those variants defective in expressing 13aa i.e. CPKO or CPΔ13aa (Fig. 2A, lanes 3 and 4 in RT-PCR panel). Since the production of STMVΔ150 is linked to HV-dependent replication [9] and the fact that expression of the 13aa plays a crucial role in STMV replication (Figs. 2A and 3B) [7]., these observations suggest that expression of the 13aa is linked to the production of STMVΔ150 by an unknown mechanism.

To test the effect of the N-terminal13aa deletion on virion assembly, agrotransformants of CPKO, CPΔ13aa and 13aa were co-expressed with agrotransformant pRP in *N. benthamiana* leaves. Plants co-infiltrated with pSTMV and pRP served as positive controls while those infiltrated only with empty vector served as negative controls. At 4 dpi, virions were purified, negatively stained and examined by Transmission Electron Microscopy (TEM). Icosahedral virions of 18 nm characteristic of STMV were recovered from leaves infiltrated with pSTMV and pRP (Fig. 2C). By contrast, no virions were detected in leaves infiltrated with any of the three CP variants (Fig. 2C). In agreement with previous findings [5], our observations further confirm that the N-terminal13aa region is required for virion assembly.

**A positively charged amino acid at position 3 is obligatory for efficient replication of STMV.** Results shown in Fig. 2A accentuate the importance of the N-terminal13aa region in STMV replication. To precisely identify the role of positively charged aa encompassing the N-terminal13aa of STMV CP on replication, an alanine residue was substituted for a desired arginine or lysine residues normally located within the 13aa N-terminal region resulting in the construction of a set of five variants of CP 13aa (i.e. 13aa/3A, 13aa/5A, 13aa/7A, 13aa/10A and 13aa/11A in Fig. 3A). Each of these CP agrotranformants was co-infiltrated with pRP into *N. benthamiana* leaves and progeny were evaluated using Northern blot and RT-PCR analyses (Fig. 3). Northern blots were probed for assessing the affect of CP and its variants on the progeny accumulation (Fig. 3B) and quantitated as described above (Figs. 3C).

First, we evaluated the affect of the N-terminal 13aa motif and five variants on plus-strand accumulation (Fig. 3B, panel i). Except for variant 13aa/3A, plus-strand accumulation for the remaining four variants was indistinguishable from that of the internal control (i.e. wt) (Fig. 3C). Secondly, we evaluated the effect of these mutants on minus-strand accumulation (Fig. 3B, panel ii). Since each of the five alanine substitutions were engineered into the genetic background of the wt 13aa construct, data was normalized against wt 13aa. Except for variant 13aa/3A, where the (-)-strand accumulation was equivalent to that of the control (i.e. 13aa), a 3-fold increase in minus-strand accumulation was observed for the remaining four variants (Fig. 3C). Taken together, these observations accentuate the significance of that having a positively charged residue at their respective positions, specifically at $3^{rd}$ position is obligatory for sustained replication.

Results of RT-PCR analysis evaluating the production of STMV Δ150 are shown in Fig. 3B. Production of STMV Δ150 was found associated with all variants except for 13aa/3A (Fig. 3B). Collectively, data encompassing replication profiles (Fig. 3B), quantitative analysis of strand asymmetry (Fig. 3C) and RT-PCR (Fig. 3B) analysis suggested that the arginine and the lysine residues at the N-terminus plays a significant role for maintaining wild type replication of the STMV RNA. More specifically a positively charged residue at the $3^{rd}$ position is intimately involved with the replication of STMV (also see below).

**Replication of full-length STMV CP variants harboring N-terminal mutations.** Next we wanted to evaluate the replication and packaging phenotypes of full-length STMV CP bearing each of the six N-terminal alanine substitution variants(Fig. 3A). Consequently, each of the five-alanine substitution variants (Fig. 3A) was engineered into the CP ORF of the full-length STMV agrotransformant (Fig. 4A) and its effect on replication, translation, virion assembly and genome packaging was evaluated. Results are shown in Figs. 4 and 5.

Northern blot hybridization results on the effect of the five-alanine mutations on STMV replication are shown in Fig. 4B and the quantitative data on the relative accumulation levels of (+) and (-)-strand progeny is shown in Fig. 4C. Consistent with the data shown in Fig. 3B, despite near wt levels of (-)-strand accumulation, a >90% reduction in the accumulation of (+)-strand was observed for variant 3A (Fig. 4B, lane 1; Fig. 4C). For the remaining four variants (ie. 5A, 7A, 10A and 11A), a 30 to 50-fold stimulation (in relation to wt) in the accumulation of (+)-strand was observed (Fig. 4B, lanes 2-5) which was contrary to the accumulation level seen with the same variants engineered in the CP 13aa construct (Fig 3). Results of RT-PCR assays used to evaluate the production of STMV Δ150 are similar to those shown in Fig. 3B, i.e. STMV Δ150 was present in total RNA preparations of all variants except 3A (Fig. 4B, bottom panel).

Results shown in Figs. 3 and 4 clearly indicate that at position 3, a positively charged amino acid residue is preferred. To substantiate this, the arginine residue at position 3 was substituted with another basic amino acid namely, a lysine residue (i.e. 3R to 3K, Fig. 4A) and its effect was examined on replication, CP synthesis and production of STMV Δ150. Results summarized in Fig. 4 (D and E) confirm the importance of a positively charged residue at position 3.

## Effect of N-terminal mutations on STMV assembly and packaging.

Western blot analysis revealed that all four other variants (5A, 7A, 10A and 11A) were competent to translate higher levels of CP than wt. Among four variants, CP production was more prominent with variant 7A than others (Fig. 4E). By contrast, CP for variant 3A was only detectable when a 10-fold excess of total protein was loaded (Fig. 4F). To verify whether these variants were competent to assemble into virions or virion like particles (VLPs), virions were purified from infiltrated *N. benthamiana* leaves. Fig. 5 summarizes the results. EM and Northern blot hybridization evaluated the physical morphology of purified virions and their packaging phenotypes, respectively. The biological nature of the purified virions was tested by an infectivity assay. Results are summarized in Fig. 5. Icosahedral virions of 18 nm were detected in wt as well as in variants 5A, 7A, 10A and 11A but not in variant 3A (Fig. 5A). Northern blot analysis of viron RNA profiles revealed interesting results. For example, only virions of wt contained full-length STMV RNA as well as truncated STMV (i.e. STMV $\Delta$150) (Fig.5B, lane 6), a characteristic feature of STMV in *N. benthamiana* [9]. By contrast, virions of all four variants contained only the truncated STMV $\Delta$150 but not full length (Fig. 5B, lanes 2-5). Presence of full length and STMV $\Delta$150 in virion preparations of wt and the absence of full length and presence of STMV $\Delta$150 in all four CP variants were further confirmed by an RT-PCR assay (Fig. 5C). In contrast to the wt control (Fig. 5D, lane 1), the inability to induce infection in *N. benthamiana* plants mechanically inoculated with virions of HV and each of the four CP variants (Fig. 5D, lanes 2-6) further confirm that CP variants are incompetent to package full length STMV RNA. Reasons for this defective packaging phenotype are considered under the Discussion.

## Evidence for a physical interaction between HV-RdRp and STMV CP.

Although replication of STMV depends on HV RdRp, results shown in Figs. 2-5 demonstrate that STMV CP, more specifically its N-terminal 13aa motif, has a regulatory role in the replication. Therefore to test whether STMV CP physically interacts with HV RdRp, we performed the following experiment. One set of *N. benthamiana* plants was infiltrated to co-express HV RdRp (i.e. pRP, Fig. 1C) and either wt STMV or one of three selected N-terminal CP variants (3A, 5A and 7A). Another set of *N. benthamiana* plants was infiltrated to co-express pRP and only 13aa tagged with FLAG (13aa$^{c\text{-}FLAG}$). Plants co-infiltrated with P19 and either pSTMV or 13aa$^{c\text{-}FLAG}$ served as controls. At 4 dpi, total proteins were extracted from infiltrated leaves and incubated with anti-126 kDa antibodies, followed by the precipitation of the complex with protein G-agarose beads and Western blotting with anti-126 kDa, CP or Anti-FLAG antibodies. Results are shown in Fig. 6A. It was observed that wt CP, variants 5A and 7A, but not 3A, were co-precipitated with HV RdRp (i.e. Anti-126 kDa) (Fig. 6A, panel ii). These results validate that a direct or indirect interaction exists between HV-RdRp and STMV CP and the arginine residue located at position 3 is obligatory in promoting this interaction. In addition, RNase A treatment did not disrupt the co-precipitation of anti-126 kDa and CP (Fig.6B), suggesting that RNA is not involved in promoting the interaction between HV RdRp and CP.

## Evidence showing interaction of STMV RNA with HV RdRp requires CP.

Northern blot results (Figs. 2-4) demonstrate the importance of CP in STMV replication while results of Co-IP assays (Fig. 6A) demonstrate that CP physically interacts with HV RdRp. Does interaction of HV RdRp with STMV RNA require CP? To answer this question, we performed a Far-Northwestern assay. Results are shown in Fig. 7 (A, B). Since variant 3A translates inefficiently *in vivo* (Fig. 4F), a Northwestern analysis was performed using *in vitro* translation products of wt and four CP variants (CPKO, 3A, 5A and 7A) and $^{32}$P-labeled STMV RNA as a binding probe. Interaction between STMV RNA and CPs of wt, variants 5A and 7A, but not with either CPKO or variant 3A, was detectable [Fig. 7A, panel (ii)]. These Northwestern analyses confirm interaction between CP and RNA while Co-IP assays (Fig. 6A) established that CP physically interacts with HV RdRp. To further evaluate whether CP is required to promote interaction between STMV RNA and HV RdRp, we performed another Far-Northwestern analysis. Results demonstrated that the presence of STMV CP enabled the interaction between HV RdRp and STMV RNA [Fig. 7B, panels (I to iii)]. Taken together these results provide evidence that CP enhances selection of STMV RNA templates by HV RdRp.

## Discussion

Viral replication is a complex process involving numerous macromolecular interactions [12]. In positive strand RNA viruses, although replication is primarily catalyzed by virus encoded RNA-dependent-RNA polymerase (RdRp), several lines of evidence suggest that CP is intimately associated with this active process to stimulate replication [12]. For example, CP-regulated viral RNA synthesis occurs via CP-RNA interactions as in the case of *Alfalfa mosaic virus* (AMV) [13,14] or CP-RdRp interactions as in the case of member viruses of the Coronaviridae [15] and rubella virus [16] or CP functioning as a RNA chaperone [11,17,18]. The results presented here demonstrate that a cooperative interaction between HV- RdRp and STMV CP (Fig. 2A), more specifically the N-terminal13aa region (Figs. 3 and 4), plays an important role in STMV replication and packaging. Mutational analysis further revealed that a positively charged residue at the third position in the N-terminal 13aa motif (Figs. 3B, 4B, 4D and 6A) promotes the physical interaction between HV-RdRp and STMV CP. These features are commonly shared with AMV [14,19] and rubella virus [16]. In AMV, minus-strand synthesis requires the viral RNA to be in a 3' pseudoknot conformation and to be free of CP; upon CP binding a conformational switch blocks minus-strand synthesis and favors plus-strand synthesis [20,21]. However, since the 3' UTR of STMV inherently mimics a TLS, a conformational switch analogous to AMV is not required to initiate minus-strand synthesis. Similar levels of minus-strand accumulation between wt and the CP defective variant CPKO (Fig. 2A, compare lanes1 and 3) suggest that in STMV minus-strand synthesis by HV RdRp is independent of the CP.

Our results exemplify that STMV CP (Fig. 2), specifically the basic N-terminal 13aa motif (Figs. 3A and 4A) actively participates in modulating (+) an/or (-)-strand synthesis. As discussed below, this active participation in STMV RNA replication depends on the genomic context of the N-terminal 13aa motif. First, the critical role played by the CP is exemplified when the replication profile of wt and mutant CPKO is compared (Fig. 2A). Since the levels of (-)-strand accumulation for wt and mutant CPKO are indistinguishable (Fig. 2A, compare lanes 1 and 3). It is reasonable to conclude that CP has no detectable effect on (-)-strand synthesis and a significant reduction in (+)- strand synthesis by mutant CPKO (Fig. 2A, lane 3) suggests CP plays a critical role in (+)-strand synthesis.

Co-expression of wt 13aa (Fig 2) along with HV RdRp stimulated both (+) and (-) strands accumulation equally (Fig 2A lane 5) whereas co-expression of ∆13aa significantly reduced (+)-strand accumulation compared to that of (-)-strand (Fig 2A compare lane 4 and 5). This suggests that the role of N-terminal 13aa motif is attributed to (+) strand synthesis. However, on mutating four of the five positively charged amino acids to alanine in the wt 13aa (i.e. 5A, 7A, 10A and 11A), the balanced accumulation of (+) and (-)-strand RNA was altered resulting in increased accumulation of (-)-strands (Fig. 3B). Interestingly, this scenario was changed when the above-mentioned four mutants were incorporated into the background of full length CP. This is exemplified by increased accumulation of (+)-strands over the (-)-strands (Fig. 4B). Collectively, these observations suggest that (+) and (-)-strand accumulation is regulated by the context of the N-terminal 13aa motif based.

Another interesting outcome of our study is the intrinsic role played by a positively charged residue at position 3 of the N-terminal 13aa motif. For example, substitution of an alanine residue for arginine at position 3 (i.e. 3R➜3A; Figs. 3A and 4A) had very little effect on (-)-strand synthesis (Fig. 3B, line 4; Fig. 4B, line 1) while plus-strand synthesis was down regulated by nearly 80% (Figs. 3B, C and 4B, C). An analogous scenario was observed in AMV. Mutational analysis of the N-terminal basic residue of AMV CP revealed that substitution of alanine for lysine at position 18 completely abolished replication [19]. As discussed below, this down-regulation of progeny accumulation is due the absence of physical interaction of CP with HV-RdRp.

Results from co-immunoprecipitation analysis revealed that STMV CP interacts with HV RdRp (Fig 6A lane 4). Although this phenomena was previously observed in other RNA viruses such as brome mosaic virus and flock House Virus (15) but not for satellite viruses. Results of this study represents first example demonstrating that STMV RNA replication requires a physical interaction between CP and its HV-RdRp. Furthermore, North-Western and Far North Western analysis (Fig. 7, A and B) accentuated that the presence of a positively charged residue at the $3^{rd}$ position is obligatory to promote CP-HV RdRp interaction.

We recently demonstrated that replication of STMV in *N. benthamiana*, but not in *N. tabacum*, generate a truncated RNA (i.e. $\Delta150$ characterized by lacking the 3' terminal 150 nt) along with the full length RNA (9). Results of this study revealed that production of $\Delta150$ is linked to HV-dependent replication since it was not detected in plants infiltrated with replication defective variants of STMV (i.e. CPKP and 3A; Figs. 3 and 4).

In general, it is widely accepted that electrostatic interactions are the driving force for virion assembly in that the positively charged domains on viral CPs neutralize most of the negative charges of the RNA[22-29]. Many *in vitro* viral self-assembly studies show that RNA packaging can be largely explained through electrostatic interactions [24]. Based on these experimental observations, we recently introduced a very simple generic model to study the spontaneous encapsulation of genome by the CP[30]. In this model, we consider RNA as a negatively charged polymer interacting attractively with positively charged CP. More specifically, we use a phenomenological model to investigate the total free energy of the virion, $F_{virion}=F_{capsid}+F_{RNA}+F_{int}$, with $F_{capsid}$ the free energy of the capsid, $F_{int}$ the free energy associated with capsid-RNA interactions and $F_{RNA}$ the free energy of RNA. Since STMV CP is positively charged, we model the capsid as a charged sphere so the term $F_{capsid}$ involves the repulsive electrostatic interaction between positive charges on the CP. In our approach, RNA is modeled as a self-avoiding flexible branched chain. Thus the term $F_{RNA}$ includes the contribution of RNA entropy, the secondary structure of RNA, and the excluded volume interaction related to the second viral coefficient. We find the optimal length of RNA and its distribution within the capsid through minimization of the free energy $F_{virion}$ explained above.

Data shown in Fig. 8 demonstrates the result of our calculations, $F_{virion}$ vs genome length. The σ in the figure corresponds to the charge density of the inner wall of the capsid. It is 0.24 electron/nm$^2$ for the capsid constructed by mutated coat proteins (5A, 7A, 10A and 11A) and it is 0.3 electron/nm$^2$ for the wt capsid. The solid black curve shows the free energy of RNA encapsulated by wt proteins while the dashed red curves correspond to that by the mutated coat proteins. Blue dots and orange squares respectively represent the encapsidation free energy of STMVΔ150 (908 nt) and wt RNA (1058 nt). As illustrated in the figure, in the case of mutated capsid, the encapsidation free energy of STMVΔ150 RNA is significantly lower than that of wt RNA ($14.28k_BT$) leading to the exclusive encapsulation of STMVΔ150 RNA. However, in the case of wt CP, the encapsulation free energy of STMVΔ150 RNA is lower than that of wt RNA by only $1.47k_BT$, explaining why both STMVΔ150 and wt RNA can be encapsidated by wt CP. As a matter of fact, while the ratio of free STMVΔ150 to wt RNA is very low (24:76) [9], the ratio of packaged STMVΔ150 to wt RNA is much larger (55:45) indicating a clear preference for encapsidation of the shorter RNA. More specifically, these calculations reveal that the electrostatic interaction between mutated capsid proteins and full length RNA is not strong enough to compactify the genome to fit in a small size capsid. In other words, due to the size of STMV capsid, the energetic cost of encapsidation of negatively charged wt RNA is very high and thus the shorter genome is encapsidated more favorably even by wt capsid proteins. While the number of charges on the wt capsid proteins is sufficient to encapsidate the full length genome, it is important to note that wt RNA gets encapsidated much less efficiently than Δ150. These theoretical results offer the basis to explain results presented in Fig. 5C.

As noted above, electrostatics clearly explains the results corresponding to variants 5A, 7A, 10A and 11A presented in Fig. 5C. However, a question that naturally arises would be why does the CP bearing the 3A mutation not encapsulate either STMVΔ150 or wt RNA? We offer the following explanation. Our observations (Fig. 5C) suggest that non-specific interactions are insufficient to drive the self-assembly of a ssRNA virus by CP. The presence of an arginine residue located at position 3 reveals the necessity of a specific interaction between RNA and CP for the successful assembly of a virion. The interaction could be due to a specific nucleation site on the CP for RNA or the specific CP-RNA interaction could alter the conformation of the CP resulting in a stronger protein-protein interaction leading to virion assembly. The experiments presented in this paper show that an amino acid residue (i.e. arginine) at a specific site (i.e. position 3) is crucial for successful packaging. Identification and disruption of contacts between particular CP residues and RNA could contribute to our knowledge of developing novel antiviral drug strategy for RNA pathogenic to pathogenic to humans and animals.

## Methods

**Agro-constructs and Agro-infiltration**. The construction and characteristic features of an agro-construct of STMV (pSTMV; Fig. 1A) have recently been described [9]. All CP variants constructed in this study are incorporated into the genetic background of pSTMV using a mega PCR approach [31]. Briefly, to construct the N-terminal CP variants (Fig. 1C) 3R➔A, 5K➔A, 7K➔A, 10R➔A, 11K➔A and 5R+7K➔5A+7A (refer to as 57A; Fig. 1C), STMV CPKO, STMV 13aa, STMV 13aa/A and STMV 13aa/B(Fig 5A), the CP gene harboring nucleotides (nt) 159 to 614 was first amplified with the following forward primers (For 3R➔A 5'GCTATGGGG<u>GCA</u>GGTAAGGTTAAACC3', 5' GCTATGGGGAGAGGT<u>GCA</u>GTTAAACC3',for 5K➔A 5' GGGGAGAGGTAAGGTT<u>GCA</u>CCAAACC 3'; for7K➔A 5' GGTAAGGTTAAACCAAAC<u>GCA</u>AAATCGACG 3'; FOR 10R➔A 5' GGTTAAACCAAAC CGT<u>GCA</u>TCGACG 3'; 5' for CPKO, 5'CTGTTTCCAGCT<u>TAG</u>GGGGAGAGGT<u>TAG</u>GTTAAACC3';for 13aa, 5'CGTAAATCGACGTAGGACAATTCGAATG 3'; for 13aa/A5' GCTATGGGG<u>GCA</u>GGTAAGGTTAAACC 3'; 13aa/B 5' CCAGCTATGGGGAGA<u>GCAGCAGCAGCAGCAGCAGCAGCA</u>TCGACGTAGGACAAT TCG 3') and a commonly shared reverse primer (5' GGCGACTT<u>GTCGAC</u>AGTTGC 3'; *Sal*I site is underlined). The resulting PCR products were gel purified and used as mega primers in a PCR reaction. A region spanning the sequence of pSTMV from nt768-614 was amplified using a forward primer (5'CGCC<u>AAGCTT</u>GCATGCCTGCAGG3'; *Hind*III site is underlined) and each mega primer for the respective variant prepared above. The resulting PCR products were digested with *Hind*III and *Sal*I and sub-cloned into a similarly treated pSTMV.  The nature of all recombinant clones was verified by DNA sequencing. Transformation of wild type and variant pSTMV agroconstructs to *Agrobacterium* strain GV3101 followed by infiltration into the abaxial side of the fully expanded *N. benthamiana* leaves was as described previously [32].

**Mechanical inoculation, progeny analysis, packaging assays and EM analysis.** Procedures used for mechanical inoculation of *N. bethamiana* with STMV variant virions in the presence of TMV U5, isolation of total RNA from agroinfiltrated leaves, purification of STMV virions, Northern blot analyses are as described previously[9,33,34] For electron microscopy analysis, purified wild type and variant STMV virions were further subjected to 10%-40% sucrose gradient centrifugation and spread on glow-discharged grids followed by negative staining with 1% uranyl acetate prior to examination with JEM 1200-EX transmission electron microscope operated at 80 keV, and images were recorded digitally with a wide-angle (top mount) BioScan 600-W 11K pixel digital camera [33].

**RT-PCR**. Virion RNA from wild type and CP variants were subjected to Poly–A tailing using *E. coli* poly A polymerase according to the manufacturer's protocol (NEB). Following inactivation of poly A polymerase by heat (95ºC for 2 to 3 minutes), first strand cDNA was synthesized using a reverse primer (5'

GGGAGGACACAGCCAACA<u>TACGTA</u>TTTTTTTTTTTTTTTTTTTTTTT 3'; *Sna*B I site is underlined) and M-MulV reverse transcriptase (New England Biolabs). The resulting product was subjected to PCR using a forward primer (5' TACGTAAACTTACCAATCAAAAG 3') and a reverse primer (5'GGGAGGACACAGCCAACA<u>TACGTA</u>3'; *Sna*B I site is underlined). PCR products were finally analyzed by agarose gel (1%) electrophoresis then visualized and photographed under UV light following ethidium bromide staining.

**Co-immunoprecipitation assay (Co-IP)**. Total protein extracts were prepared from healthy and agroinfiltrated *N. benthamiana* leaves as described previously[35]. Isolated proteins were incubated with anti-126 kDa protein antibody at 1:100 dilution (kindly provided by Rick Nelson) for 8 h at 4°C. Then, a 30-$\mu$l aliquot of protein G-agarose beads (Santa Cruz Biotechnology, USA) was added to each tube, followed by incubation for 2 h at room temperature. The immune complexes were then precipitated by centrifugation for 1 min at 10,000 x g and washed three times in 1 ml of phosphate-buffered saline (0.1 M NaCl, 90 mM sodium phosphate [pH 7.0]). For RNase A treatment, the precipitated proteins were treated with RNase A (50 $\mu$g/ ml) for 2 h at 25°C[31]. The resulting precipitated proteins were eluted from the beads by boiling in SDS-PAGE sample buffer for 3 min. Equal volumes of protein samples were analyzed by SDS-PAGE, followed by immuno blot analysis with anti-126-kDa and anti-STMV CP antibodies.

**In-vitro translation of CP**. *N. benthamiana* leaves were co-infiltrated with agrotransformants of STMV wt or CP variants 3A or 5A or 7A and TBSV P19, a suppressor of RNA silencing [36]. Total RNA isolated from each of these infiltrated leaves was used to produce their respective CPs using a wheat germ *in vitro* translation kit (Promega Corporation) and subjected to Western, Northwestern and Far-Northwestern blot analyses.

**Western, Northwestern and Far-Northwestern analyses**. Western blot analysis was performed as described previously [9]. For Northwestern analysis [14], wild type or desired variant CPs of STMV were resolved by 10% or 12% SDS-PAGE and transferred to a nitrocellulose membrane. Proteins immobilized on the nitrocellulose membrane were re-natured overnight in a buffer containing 15 mM HEPES (pH 8.0), 10 mM KCl, 10% glycerol, and 1 mM dithiothreitol at 4°C. Membranes were then hybridized with $^{32}$P-STMV RNA for 1 h at room temperature in re-naturing buffer containing 2 mg/ml of yeast tRNA. The membrane was washed twice with re-naturing buffer at room temperature to remove any unbound RNA, followed by autoradiography. For Far-Northwestern analysis [14], after performing Northwestern analysis as described above, hybridization of the membrane was performed with a mixture containing $^{32}$P-labelled STMV RNA transcripts and *in vitro* translated STMV CP.

## Acknowledgments

We wish thank John Lindbo for the pJL 36 construct, Barbara Baker for the pRP construct and Richard Nelson for anti-126kDa RdRp antibody. This work is supported in part by an RSAP grant from UCR (A.L.N.R and D.M) and the National Science Foundation Grant No. DMR-1310687 (R.Z).

## Author information

### Affiliations

**Department of Plant Pathology & Microbiology, University of California, Riverside, CA 92521-0122**

Venkatesh Sivanandam, Deborah Mathews, A.L.N. Rao

**Department of Physics & Astronomy, University of California, Riverside, CA 92521, USA**

Gonca Erdemci-Tandogan, Roya Zandi

**Department of Chemistry & Biochemistry, University of California, Los Angeles, CA, 90095, USA**

Rees Garmann

## Contributions

ALNR and RZ designed the project and experiments; VS, RG and GET performed the experiments; DB contributed reagents/ materials; ALNR and RZ analyzed the final data and wrote the manuscript. All authors read and approved the final version of the manuscript.

## Competing interests

The authors declare no competing financial interests.

## Corresponding author

Correspondence to A. L. N. Rao

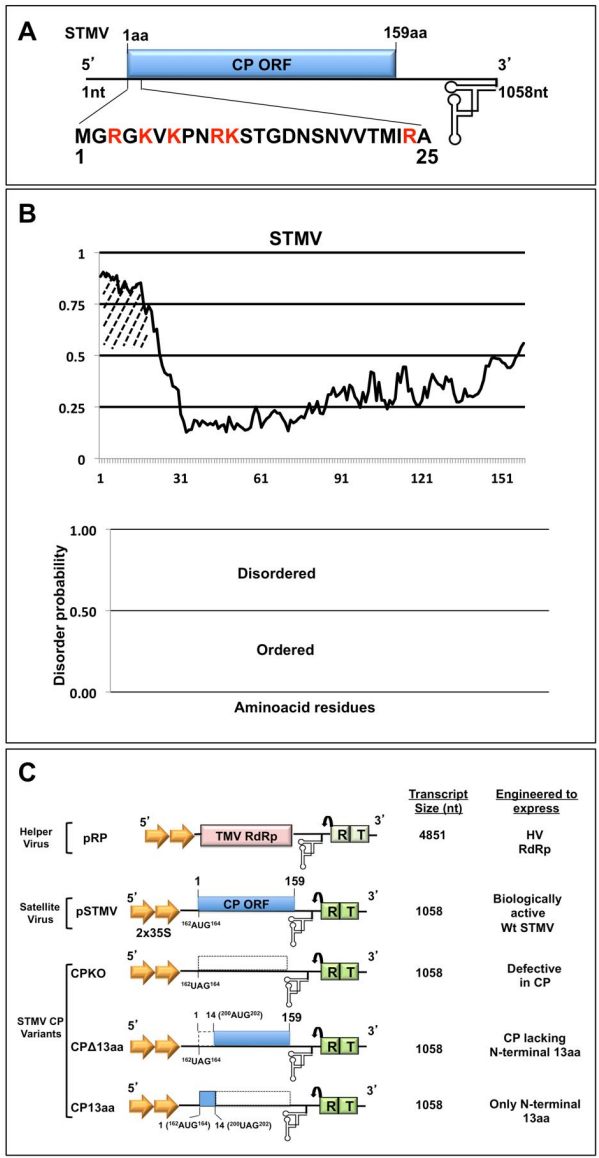

**Figure 1. Genome organization of STMV.** A) Genome organization of *Satellite tobacco mosaic virus* (STMV) highlighting the 25 N-terminal amino acids (aa) of its coat protein ORF. The positively charged arginine (R) and lysine (K) residues are highlighted in red. B) Graphic representation of the disordered probability of the 159 aa residues of STMV CP was by DisProt analysis. The hatched region indicates a previously identified RNA-binding domain (see text for details). In the bottom panel, aa having a score equal or above 0.5 and below 0.5 are considered to be disordered and ordered, respectively.

C) T-DNA based agro-constructs of STMV and its CP derivatives. Characteristic features of agroconstructs pRP and pSTMV were described previously [9]. CPKO, an STMV CP variant is characterized by having a $^{162}$UAG$^{164}$ stop codon in the place of an$^{162}$AUG$^{164}$ start codon. CP variant CPΔ13aa is engineered to express CP lacking the N-terminal 13aa motif due to the presence of a stop codon between nt 162-164 ($^{162}$UAG$^{164)}$ and a start codon between nt 200-202($^{200}$AUG$^{202}$). CP variant CP13aa is engineered to express only the N-terminal 13aa motif by having a stop codon at amino acid position 14 located between nt 200-202 ($^{200}$UAG$^{202}$). Each agroconstruct contains in sequential order (L to R), double *Cauliflower mosaic virus* (CaMV) 35S promoters (indicated by arrowheads), a ribozyme (R, indicated by a bent arrow) derived from *Tobacco ring spot virus* (TRSV) and a NOS terminator (T). The size in nucleotides (nt) of ectopically expressed RNA transcripts and the expected nature of proteins in each case are shown to the right. The numbers shown in parentheses are the numbers of non-viral RNA nucleotides left after self-cleavage by the ribozyme (R).

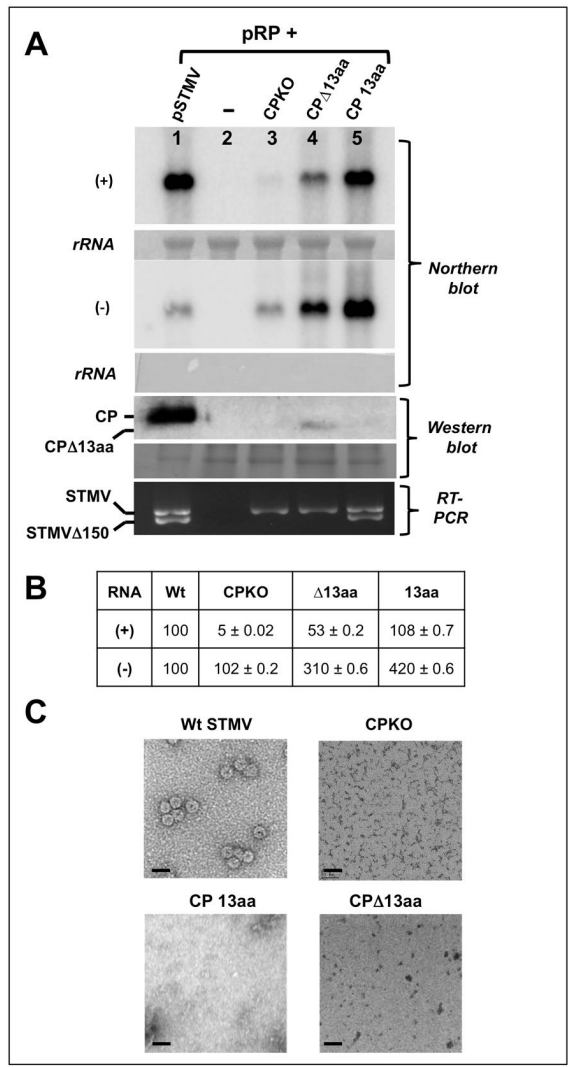

**Figure 2. STMV progeny analysis**. A) STMV plus (+) and minus (-) strand progeny RNA and CP were respectively probed by Northern and Western blot analysis. RT-PCR assay of wt STMV and the indicated CP variants was performed using total RNA isolated from leaves harvested at 4 dpi. B) Quantitative analysis of (+) and (-) progeny RNA of wt STMV and each CP variant was performed by scanning the autoradiographs using a Typhoon Phosphoimage scanner (Model 9410, GE Health Care) followed by ImageQ software analysis. Data was quantitated and compared internally with each levels of (+) and (-)-strands to that of wild type. Conditions for Northern and Western blots and RT-PCR analysis are as described previously [9].The positions of full length and truncated CP, full length and truncated STMV RNA are indicated on the left. C) Virion analysis. The physical morphology of negatively stained purified virions of wt STMV (left panel) and its CP variants (right panel) was examined by Electron microscopy [9]. Bar=50 nm.

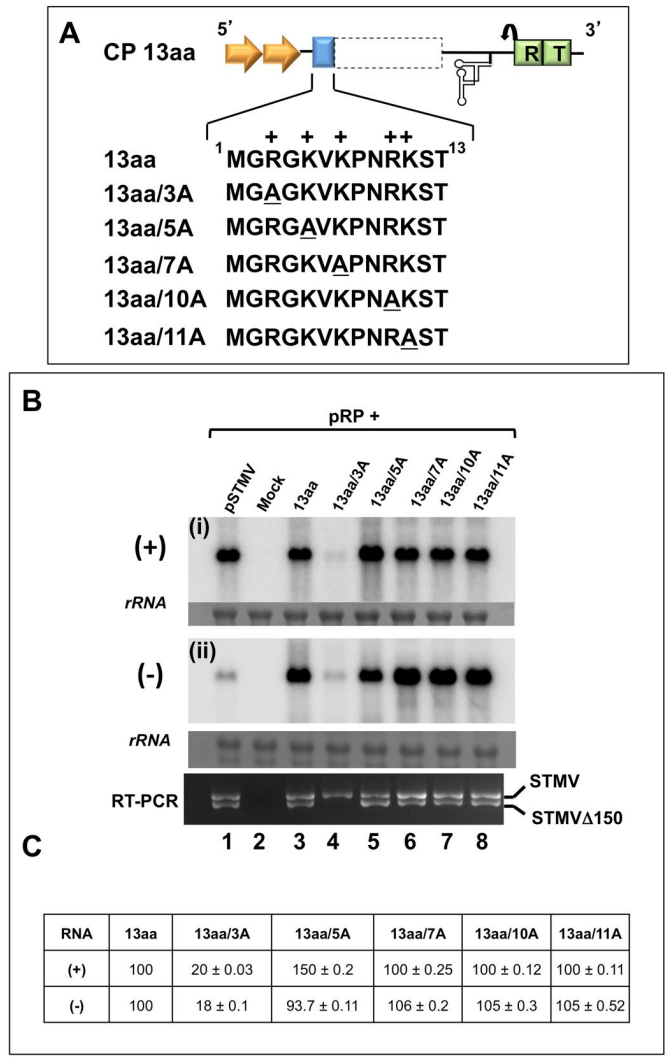

**Figure 3. Mutational analysis of the N-terminal 13aa motif of STMV CP. (**A)
Characteristic features of the agroconstruct of CP 13aa is shown under Fig. 1C. The N-
terminal 13aa sequence of the CP is indicated below the line diagram. In the clone
13aa, the location of the positively charged amino acids are indicated by a + sign. In
each variant clone the location of the engineered alanine mutation is indicated by an
underline. The nomenclature used to designate each variant is exemplified by 13aa/XA,
where X is characterized by an alanine (A) substitution for arginine (R, at positions 3

and 10) or lysine (K at positions 5, 7 and 11). (B) Progeny analysis of N-terminal 13aa motif variants by Northern blot and RT-PCR. Duplicate blots containing total RNA of *N. benthamiana* leaves infiltrated with the indicated samples (lanes 1 through 8) were hybridized with a $^{32}$P-labelled STMV probe to detect either plus (panel i) or minus (panel ii) sense RNA. Ribosomal RNA (rRNA) represents the loading control. Conditions used for Northern blot hybridization and RT-PCR is as described under the Fig. 2A legend. Positions of STMV and STMV$\Delta$150 are indicated to the right. C) Progeny RNA quantification. Northern blots shown in panel B were scanned using a Phosphoimager. Since each mutation was engineered into the genetic background of 13aa, the absolute values of accumulated (+) and (-)-strand progeny RNA. for each mutant were compared internally to the 13aa.

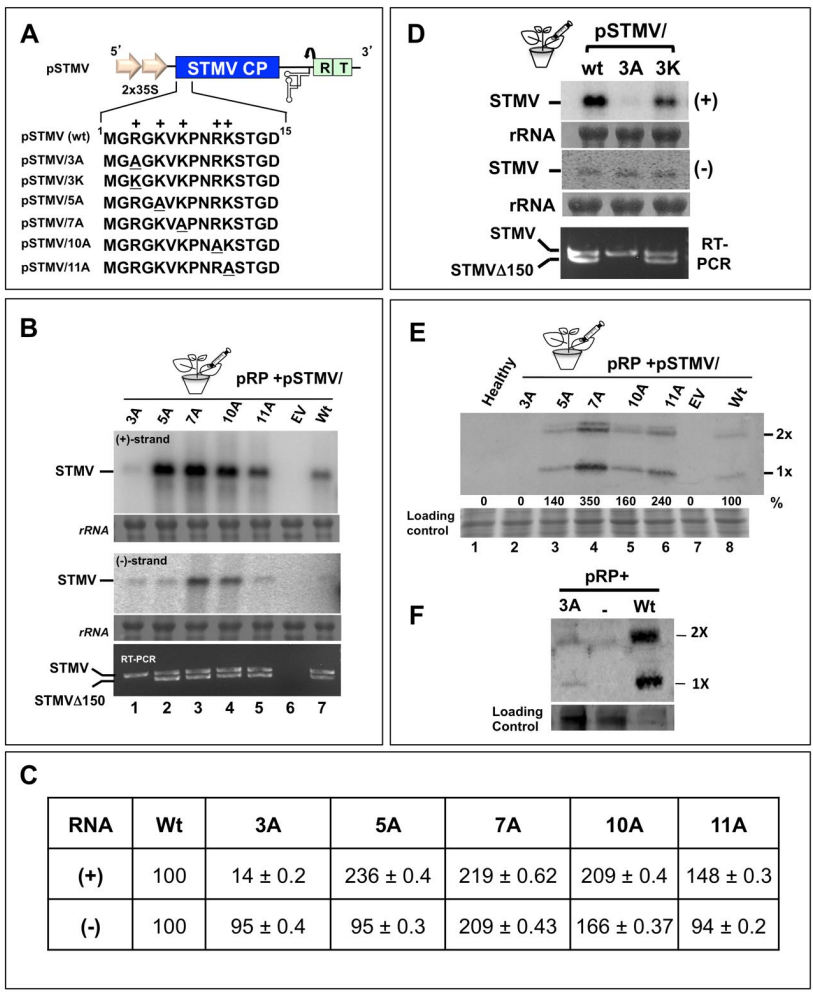

**Figure 4  Replication of full-length STMV RNA bearing mutations in the N-terminal 13aa motif**. (A) Mutations engineered into the N-terminal 13aa motif of CP. Positively charged amino acid residues are indicated by a + sign.  Alanine or lysine substitutions at designated positions are underlined. Nomenclature used to designate each mutation is the same as described under the Fig. 3 legend. (B-E) Progeny analysis. Total RNA and protein samples were isolated from leaves harvested at 4 dpi. Panels B and D summarize Northern blot analysis of (+) and (-)-strand accumulation for the indicated

variants and RT-PCR analysis. Panel C summarizes the absolute values of accumulated (+) and (-)-strand progeny RNA. (E) Western blot analysis of CP expression for indicated variants. Following co-infiltration of pRP and each variant agrotransformant progeny were analyzed by Northern blot hybridization, RT-PCR, Western blot analysis and the absolute values of accumulated (+) and (-)-strand progeny RNA.as described under the Figs. 2 and 3 legends.  (F) Detection of CP 3A by Western blot analysis using a 10-fold excess of the indicated protein samples. Positions of STMV and STMV$\Delta$150 are indicated to the left in panels B and D. Ribosomal RNA (rRNA) represents loading controls for Northern blots. Position of monomeric (1x) and dimeric (2x) forms of STMV CP is indicated to the right in panel E. Accumulation of (+) and (-)-strand progeny was quantitated as described under Fig. 2 legend.

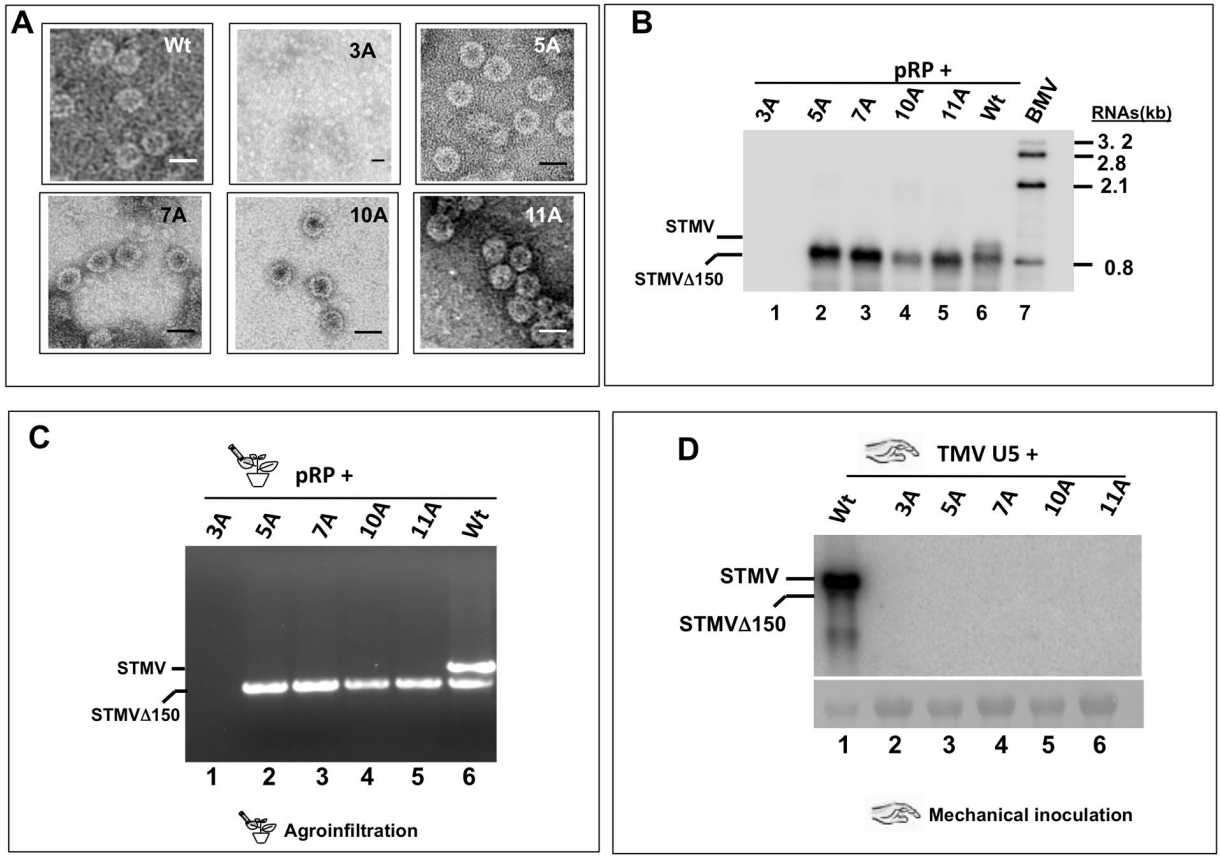

**Figure 5. Electron microscopy and packaging phenotypes of N-terminal variants.**

(A) Electron microscopic images of virions purified indicated samples. Bar= 20 nm. (B) Virion RNA analysis of indicated samples (lanes 1 through 6) by Northern blot hybridization. BMV RNA (lane 7) was used as a size marker. (C) RT-PCR analysis of virion RNA recovered from *N. benthamiana* plants co-infiltrated with pRP and either the indicated variant inocula (lanes 1 through 5) or wt STMV (lane 6). (D) Biological assay in *N. benthamiana* plants mechanically inoculated with helper virus TMV-U5 and virion RNA of wt and the indicated STMV variant samples.

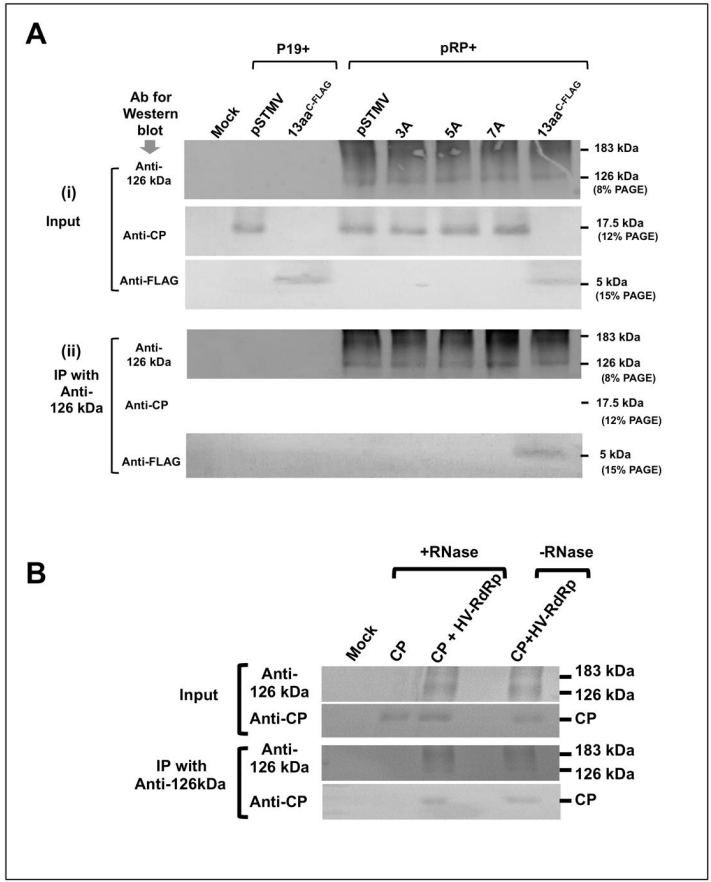

**Figure 6. Co-immunoprecipiation assays**. (A) HV RdRp (pRP) was co-expressed with either wt STMV CP (wt) or its indicated variants in *N. benthamiana* leaves by agroinfiltration. (i) Expression of HV RdRp, STMV CP and its variants in total protein extracts was confirmed by Western blotting using anti-126kDa and anti-STMV CP antibodies, respectively. Total protein extracts from leaves infiltrated with the empty vector (mock) or CP only were used as controls. (ii) For co-immunoprecipitation, total protein extracts of each sample were incubated first with anti-126-kDa antibody followed

by complex precipitation with protein-G agarose. The resulting co-immunoprecipitated products were subjected to Western blotting using anti-126 kDa and anti-STMV CP antibodies. (B) STMV CP (CP) was expressed either independently or co-expressed with HV RdRp in *N. benthamiana* leaves by agroinfiltration. Total protein extracts were divided into two batches: one batch was treated with RNase (+RNase) and the other remained untreated (-RNase). Co-immunopreciptation with anti-126 kDa followed by Western blot analysis was as described above.

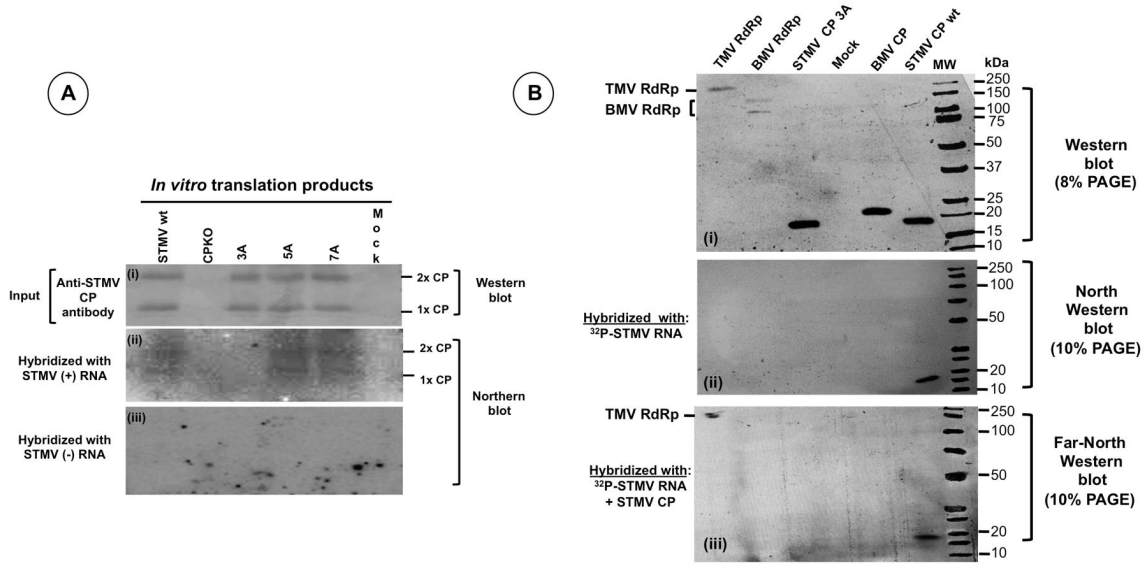

**Figure 7. Northwestern and Far Northwestern blot assays.** A) RNA-CP interaction detected by Northwestern blot analysis: Triplicate blots containing in vitro translation products of indicated samples were generated following 12% SD-PAGE analysis. The first blot (i) was subjected to Western blot analysis using anti-STMV CP antibody while the second (ii) and third (iii) blots were respectively subjected to hybridization with $^{32}$P-labeled STMV (+) and (-)-strand riboprobes and visualized using autoradiography. The positions of monomeric (1x) and dimeric (2x) forms of STMV CP are indicated to the right. B). Requirement of CP to promote interaction between HV RdRp and STMV RNA was detected by Far Northwestern blot analysis: HV RdRp (183 kDa) or, BMV RdRp (109 and 94 kDa) and BMV CP (19 kDa) (negative controls) and Wt and 3A variant of STMV CP (17.5 kDa) were fractionated by either 8% (panel i) or 10% SDS-PAGE (panels ii and iii), transferred to a nitrocellulose membrane and successively denatured and re-natured prior to hybridizing with[32]P-labeled STMV RNA probe only (ii) or radiolabelled probe and STMV CP (iii). Western blot shown in panel (i) represents input control analyzed using a mixture of antibodies raised against TMV-RdRp, BMV-RdRp, BMV CP and STMV CP antibodies.

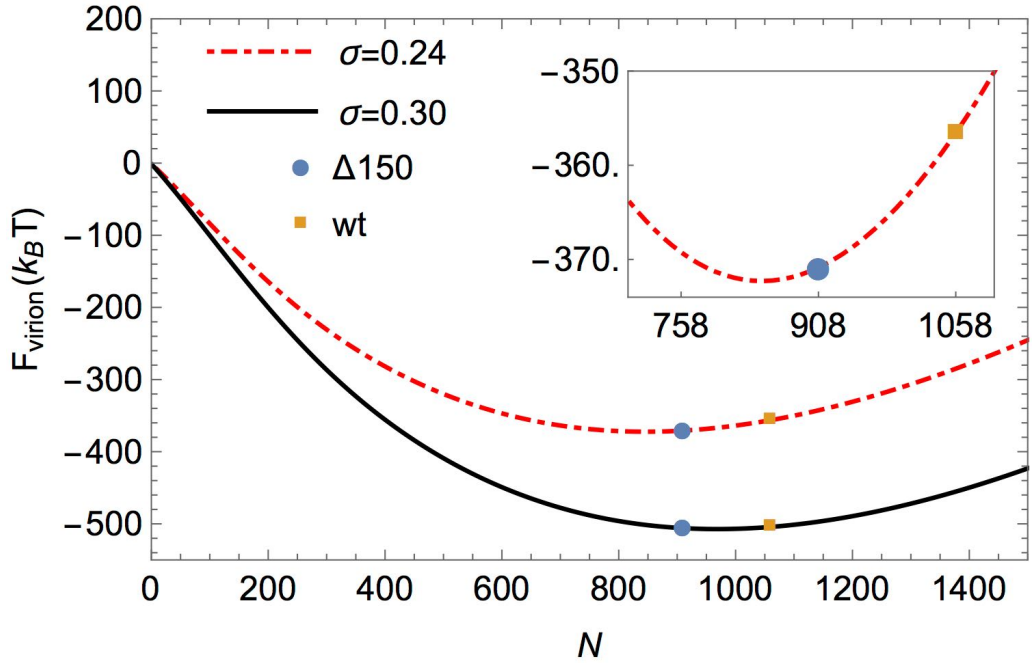

**Figure 8. Free energy of assembly.** Encapsidation free energy vs the genome length for wt CPs (solid black curve) and mutated CPs (dashed red curve). The blue dots correspond to the encapsidation free energy of STMVΔ150 RNA and the orange squares to wt STMV RNA. Energies are in units of thermal energy $k_BT$. See the text and Ref [30] for further details of the model.